\documentclass[11pt]{article}

\usepackage{amsmath}
\usepackage{latexsym}
\usepackage{amssymb}

\numberwithin{equation}{section}
\title{A Model of Anthropic Reasoning, Addressing the Dark to Ordinary Matter Coincidence
\footnote{Solicited article for the book ``Universe or Multiverse?'', ed. B. Carr, based in part on a talk at the Anthropic Workshop, Cambridge University, June 2001.}}
\author{Frank Wilczek}
\begin{document}
\maketitle

\begin{abstract}
If inflation occurs after the breaking of Peccei-Quinn symmetry then large values of the breaking scale $F$, which are forbidden in conventional axion
cosmology, are permitted, provided that we inhabit a region of the Multiverse where the initial misalignment is small. 
Regions having approximately this initial misalignment may occupy a small volume of the Multiverse,
but they contain a large fraction of potential observers.   This scenario has many consequences, including a possible explanation of the approximate equality of dark and baryon matter densities.
\end{abstract}


\section{Methodology and Anthropic Reasoning}

There are good reasons to view attempts to deduce basic laws of Matter from the existence of Mind with skepticism \cite{basic}.  Above all, it seems gratuitous.  
Physicists have done very well indeed at understanding matter
on its own terms, without reference to Mind.  We have found that the governing principles take the form of abstract mathematical equations of universal validity, which refer only to
entities -- quantum fields -- that clearly don't have minds of their own.   Working chemists and biologists for the most part are committed to the program of understanding how minds work
under the assumption that it will turn out to involve complex orchestration of the building blocks that physics describes \cite{crick}; 
and while this program is by no means complete it hasn't encountered any show-stopper, and it is supporting steady advances over a wide front.  
Computer scientists have made it plausible that the essence of mind is to be found in the operation of algorithms that in principle could be realized
within radically different physical embodiments (cells, transistors, Tinkertoys) and in no way rely on the detailed structure of physical law \cite{baum}.   To put it shortly, the emergence of Mind
doesn't seem to be the sort of thing we'd like to postulate and use as a basic explanatory principle.  Rather it is 
something we'd like to understand and explain by building up from simpler phenomena.   

So there is a heavy burden to justify use of anthropic reasoning in basic physics.  And yet there are, it seems to me, limited, specific circumstances under which such reasoning
can be correct, unavoidable, 
and clearly appropriate.  

Here is a simple, but I think instructive and far from trivial example.   Why is Earth at the distance it is from the Sun?  At one time (for example, to Kepler) the size
and shape of the Solar System -- as yet not clearly distinguished from the Cosmos as a whole -- might have seemed like a major question for physics, that one might hope would have a unique answer
closely related to basic principles.   Now, of course, the question appears quite different.  We know that the Universe contains many broadly similar systems with planets orbiting around stars.  We
know that such systems come in various sizes and shapes, and that their structure depends sensitively on details of the complicated conditions under which they formed.   
We can be confident of all these assertions 
because they emerge from a rich background involving astronomical observations, the success of Newtonian mechanics, and modern developments in cosmology and chaos theory.  
Given two key features: many independent realizations, effectively random variation over the realizations -- we cannot address our question in the context of planets in general, or
universal laws.  If we are going to address it at all, we have to refer specifically to Earth; and what makes Earth special, in this context, is that it's where we, the question-askers, find ourselves.
Once we accept this starting point we can go on to have an edifying discourse about why life would be difficult if the distance from the Earth to the Sun were quite different.   
We can even imagine that normal, testable scientific predictions will emerge from this discourse, about where we will find life in other planetary systems 
-- or even elsewhere in our own.    

A psychological weakness of this example that we've come a long way since Kepler's time, and it is hard to put ourselves back in the frame of mind to regard the Earth-Sun distance as a
serious question for physics; but however we regard the question, it seems clear that the final step in a serious answer must involve anthropic reasoning.   

The main thing I want to do here is to demonstrate that there is a choice of assumptions that, while somewhat speculative, lie well within the main stream of
present-day ideas about basic physics and cosmology, which leads to a situation whose logical structure is quite similar to this simple example, but where the question that gets addressed by anthropic reasoning
is one that is open, topical, and widely believed to be basic.  I speak of the question of the ratio $r$ of dark matter to baryon matter in cosmology.  
To be specific, I will demonstrate, given specified, reasonably conventional  \emph{physical\/} assumptions: 
that $r$ varies over the ensemble of effective homogeneous Universes within a spatially gigantic Multiverse that is inhomogeneous on superhorizon scales; that its variation is random (with a well-characterized, 
non-singular measure); and -- an important refinement --  that it varies 
essentially independently  of other parameters.   If we measure the probability by volume, we find that overly large values of $r$ are most probable, but if we measure by number of potential observers, 
this conclusion is
changed.  It appears instead that the observed value of $r$ is at least qualitatively, and perhaps semi-quantitatively, in accord with what these ideas suggest.  That is intriguing,
because most according to alternative, more conventional ideas about $r$ the numerator and denominator arise from widely different physical causes and depend upon widely
different parameters, so it has appeared as something of a mystery why the observed value is near unity.   There are several additional implications of the assumptions 
that can be explored in future experiments.

The possibility of avoiding the bound that arises in conventional axion cosmology by having a small 
misalignment after inflation was already mentioned in the earliest papers on axion cosmology (i.e., the first paper in \cite{acosmo}).   It was exploited in the context of a specific inflationary model in \cite{pi}, where the variation of the dark matter to baryon matter ratio over different regions of the universe was noted.  Anthropic considerations were brought into the discussion in \cite{linde}.   Some constraints on the scenario, due to the axion field supplying an additional source of fluctuations, were derived in \cite{turnerwilczek}.

\section{Conventional and Unconventional Axion Cosmology}

I will now very briefly review the relevant aspects of axion physics and cosmology. 

\subsection{Axion Physics} 

QCD is well established as the basic (and fundamental) theory of the strong interaction \cite{shifman}.  
When we combine QCD with the electroweak interactions, however,  a subtle but I believe quite profound puzzle arises, as follows \cite{axion}.   
The general principles that define QCD -- relativistic quantum field theory and gauge symmetry -- specify its structure extremely 
tightly.  The continuously adjustable parameters of the theory are a single overall coupling constant, a mass for each quark, and one other much more obscure parameter, the so-called $\theta$ parameter.
Mountains of data described by QCD precisely determine and vastly overdetermine the coupling and masses, and the description it affords is more than satisfactory.   

Amidst this otherwise splendid party,  
the $\theta$ parameter appears as an empty chair, an invited guest whose absence is cause for concern.   The $\theta$ parameter is a periodic variable whose possible values range from $0$ to $2\pi$.
It specifies the phase $e^{i\theta}$ which accompanies the occurrence of special topological features in the color gluon field.  One measure of its subtlety is that $\theta$ cannot 
be detected in perturbation theory.   Under space inversion P or time reversal T $\theta$  changes sign, so that for $\theta \neq 0, \pi$ these symmetries are violated. 
There are very stringent experimental constraints on P or (especially) T violation in the strong interaction, especially from the upper limit on the neutron's electric dipole moment.  They indicate
$|\theta | \leq 10^{-8}$.  (The possibility that $\theta$ is near $\pi$ requires separate consideration, but is excluded on other grounds.)  If we were to regard QCD in isolation, we could simply impose P or T 
symmetry, thus naturally enforcing $\theta=0$.   But in a complete world-theory we must acknowledge that P and T are not exact symmetries of the world, and we cannot invoke them to justify  $\theta=0$.
We must look for another way of explaining the smallness of $\theta$.   

Peccei and Quinn (PQ)  introduced the idea that there is a special sort of approximate symmetry, valid asymptotically at short distances, that could be used to address this challenge.  
The PQ symmetry transformations allow
translations of $\theta$.  If it were exact all values of $\theta$ would be physically equivalent, and of course they would all preserve P and T for the strong interaction (some field redefinitions might be required
to make the symmetries manifest).   In reality PQ symmetry must be spontaneously broken, since in its unbroken form it is inconsistent with non-zero quark masses.   To capture this
dynamics, we introduce a complex scalar order parameter field $\phi$.  The average value  $\langle \phi \rangle$ will vanish when PQ symmetry is unbroken, but will take the form 
$\langle \phi \rangle = F e^{i\theta } \equiv F e^{i \frac{a}{F}}$ in the unbroken phase.  With this definition, the scalar field $a$ has its kinetic energy term, inherited from that of $\phi$, 
normalized in the canonical way.   
Furthermore PQ symmetry is not exact, but only asymptotic, even before its spontaneous breakdown.  The potential for $\phi$ is presumably of the general form $(|\phi |^{2} - F^{2})^{2}$ in the amplitude direction,
but depends on the phase only through nonperturbative effects in QCD, in roughly the form $(1-\cos \theta)\Lambda^{4}$, where $\Lambda \sim 200$  MeV is roughly the QCD scale,
here assumed to be $\ll F$.  The PQ symmetry is responsible for this structure.  
There is a difference in energy densities of order $\Lambda^{4}$ as one varies over the range of $\theta$,.  The minimum energy occurs very near
$\theta =0$.   The scalar field $a$ will tend to relax to zero, thus rendering $\theta =0$ and solving our puzzle.  

The field $a$ introduced in this way is called the axion field, and of course its quanta are called axions.  
The phenomenology of axions is essentially controlled by the parameter $F$, which specifies the amplitude of the condensate.  $F$ has dimensions of mass.  The mass 
$m_{a}= \Lambda^{2}/F$ of the axion and the strength of its basic couplings
to matter are both proportional to $1/F$.   Various laboratory and phenomenological constrains appear to require $F\ge 10^{9}$ GeV; the axion must be both extremely light and extremely weakly coupled.  

\subsection{Reference Axion Cosmology}

Now let us consider the cosmological implications \cite{acosmo}.  Peccei-Quinn symmetry is unbroken at temperatures $T\gg F$.   When the symmetry breaks the initial value of the phase, that is $e^{ia/F}$, 
is random beyond
the then-current particle horizon scale.   One can analyze the fate of these fluctuations by solving the equations for a scalar field in an expanding universe.  The only unusual feature is that the effective mass of the
axion field depends on temperature.   The axion mass is very small for $T \gg \Lambda$, even relative to its zero-temperature value, 
because the nonperturbative QCD effects that generate it involve coherent gluon field fluctuations (instantons) which are suppressed at high temperature.   It saturates, of course, for $T \ll \Lambda$. 
The full temperature dependence of the mass can be pretty reliably
estimated, although the necessary calculations are technically demanding.   

From standard treatments of scalar fields in an expanding universe, we learn that there is an effective cosmic viscosity, which keeps the field frozen so long as the expansion parameter is large compared to the mass,
$H\equiv \dot{R}/R \gg m$.  In the opposite limit $H\ll m$ the field undergoes lightly damped oscillations, which result in an energy density that decays as $\rho \propto 1/R^{3}$.  
At intermediate times there is a period of quasi-adiabatic damping.   This damping has a consequence that is very important for the present discussion, namely that the final mass density, 
normalized to ambient $T^{3}$, varies roughly proportional to $F \theta^{2}$.  
The qualitative feature, that the final density decreases with decreasing $F$, may appear paradoxical, since the axions are getting heavier, but
it is not hard to understand heuristically.   
For smaller values of $F$, corresponding to larger mass, the temperature at which the axion field begins to feels the effect of cosmic viscosity sets in earlier, and there are more damping cycles; while
the initial energy density depends only on the mismatch angle $\theta$, and is independent of $F$.
The time-oscillating field at  can be interpreted as 
pressureless matter, or dust (note that spatial inhomogeneities on small scales, which would provide pressure, begin to get damped as they enter the horizon).  
In simple words, we can say that the initial misalignment in the axion field, compared to what later turns out to be the favored value, relaxes by emission of axions in
a very cold coherent state, or Bose-Einstein condensate.  It is \emph{not} in thermal equilibrium with ordinary matter; the interactions are far too weak to enforce that equilibrium.   

If we ignore the possibility of inflation, then for the large values of $F$ of interest the horizon scale at the Peccei-Quinn transition at $T\approx F$ corresponds to a spatial region today that is negligibly small on 
cosmological scales.  Thus in calculating the axion density we are justified in performing an average over the initial mismatch angle.   This allows us to calculate 
a unique prediction for the density, given the microscopic model.  The result of the calculation is usually quoted in the form 
\begin{equation}\label{conventionalDensity}
\rho_{\rm axion}/\rho_{\rm dark} \approx F/(10^{12}~ \rm{GeV})
\end{equation}
 where $\rho_{\rm dark}$ is the dark energy.   In this way we would deduce that axions form a good dark-mater 
candidate for $F \sim  10^{12}~ \rm{GeV}$, and that larger values of $F$ are forbidden.  

These conclusions are unchanged if we allow for the possibility that an epoch of inflation preceded the Peccei-Quinn transition.  

\subsection{Alternative Axion Cosmology} 

Things are very different, however, if inflation occurs after the Peccei-Quinn transition \cite{heisVol}.  
For then the effective Universe accessible to present-day observation, instead of containing many horizon-volumes from
the time of the PQ transition, is contained well within just one.   It is therefore not appropriate to average over the initial mismatch angle.  We have to restore it as a \emph{contingent universal constant}.   That is,
it is a pure number that characterizes the observable universe as a whole, but which clearly cannot be determined from any more basic quantities, even in principle -- indeed, 
it's a different number elsewhere in the Multiverse!

In that case it is appropriate to use a different form of Equation \ref{conventionalDensity}, viz.
\begin{equation}\label{unconventionalDensity}
r \equiv \rho_{\rm axion}/ \rho_{\rm baryon} \approx 12 \frac{F}{10^{12} ~\rm{GeV} } \sin^{2} (\theta/2)
\end{equation}
This differs from the earlier form in two simple but profound respects.  First, I have normalized the axion density relative to baryon density rather than dark matter density.  
This change is completely trivial at a numerical level, of course.  (For concreteness, I've taken 
$\rho_{\rm dark}/\rho_{\rm baryon} =6$.)  It reflects, however, two important ideas: that changes in the mismatch angle $\theta$ do not significantly affect baryogenesis, 
so that the baryon density is
a fixed proportion of the photon density at high temperature, and provides an appropriate gauge for measuring the aspects of the cosmic environment apart from $F$ and $\theta$. This is true in many but 
perhaps not all plausible models of baryogenesis.   Second, I have reinstated the $\theta$ dependence.   The exact formula for this dependence is more complicated, 
but \ref{unconventionalDensity} has the correct qualitative features.  

In this alternative axion cosmology, values of $F \ge 10^{12}$ GeV are no longer necessarily inconsistent with existing observations.   
An ``over-large'' value of $F$ can be compensated by a small value of the initial mismatch $\theta$.

\section{Application of Anthropic Reasoning}

In the alternative axion cosmology $r$, through its dependence on the initial mismatch $\theta$, becomes a contingent universal constant.   
Furthermore it varies in a statistically well-categorized manner over the Multiverse; 
its variation can be considered in isolation from possible changes in other universal constants; and it has significant impact upon the possible emergence of intelligent observers.   
Altogether it appears to be an ideally
favorable case for the application of anthropic reasoning.  

Since in practice we only get to sample one effective universe, there is no question of checking statements about the probability distribution of effective universes by normal sampling methods.   
The best we can do is to calculate the probability that the outcome fits what we observe, given some measure.   
In our problem, one possible measure that suggests itself is simply unit weight per unit volume within the Multiverse, corresponding to the question: 
What does an average place look like?
Another possible measure is unit weight per unit observer within 
the Multiverse, corresponding to the question: What does an average observer observe?  The first (measure V) is quite straightforward, 
in our immediate case, while the second (measure A, for anthropic) involves
challenging issues, both practical and 
conceptual.    Can we really tell which parameters support the emergence of observers, much less calculate how many?  Do vastly more observers later count as much 
as relatively few today?\footnote{This dynamic question, it seems to me, is especially relevant to anthropic reasoning about the dark energy.  Universes with a smaller value of the effective
cosmological term can support 
intelligent life for 
longer, and plausibly populations of intelligent life once established grow exponentially, ... .  It arises even for measure V: should we take spatial volume, space-time volume, or something else?}  
Should we really try to estimate the number of intelligent entities with distinct ``selfs'' who actually form the notions of dark matter and baryons and measure $r$ -- or what?  

Here I will briefly indicate a few key issues and tentative conclusions.  A more definitive treatment is in preparation \cite{ribnick}.  

First let's consider the situation with respect to measure V.   If we define $F_R \sim 10^{12}$ GeV to be the value of $F$ that leads to the observed dark matter density in the reference cosmology, 
then the probability to observe less than or equal to the density we do is, taking the $\sin^{2} (\theta /2) $ dependence literally,
 $L \equiv  2 \sin^{-1} \sqrt {\frac{F_{R}}{2F}} / \pi$; while the probability to see more is of course one minus this. 
Note that any $F\ge F_{R}/2$  is allowed at some level, so that $F$ could even be slightly smaller than $F_{R}$.    
We might claim victory, following measure V, if neither of these probabilities is terribly small.     For $F/F_{R} = 10^{2}, 10^{4}, 10^{6}$, the latter two 
roughly representing unification and Planck scales respectively,  we
find $L =.045, .0045, .00045$.   Viewed this way, really large values of $F$ look unlikely.  

Things appear quite different from the perspective of measure A.    
For a first pass, I will suppose that the number of observers is proportional to the number of baryons.  In the relevant part of universal history,
when the cosmological term is subdominant or nearly so, the baryon density at a fixed Hubble parameter -- 
or, to an adequate approximation,  fixed age of the Universe --  depends on $r$ as $\rho_{b}/(\rho_{a}+ \rho_{b}) = 1/(1+r)$.  
Using Equation \ref{unconventionalDensity}, then, the probability that $r$ is equal to or less than $s$, according to measure A, is given by
\begin{equation}
L(s, u) = \frac {\int^{w}_{0} 1/(1 + 12u \sin^{2} \phi) d \phi}{\int^{\pi/2}_{0} 1/(1 + 12u \sin^{2} \phi) d \phi}
\end{equation}
with $w\equiv \arcsin \sqrt \frac{s}{12u}, u\equiv F/F_{R}$.   Half the probability is covered by $r \le 1$, 
but there is plenty of weight around $r=6$, even for very large values of $F$.  The probability
that $r$ lies between 2 and 10 is very nearly 20\%, whether $u$ is 10 or $10^{6}$!

Both very large and very small values of $r$ may not be smart places to live \cite{ribnick}.  At large $r$ it becomes difficult to make stars: in these baryon-poor universes
the largest objects that cool and fragment, as opposed to relaxing into diffuse virial clouds,
are too small to make stars efficiently.   At small $r$ we have baryon-dominated universes, and we get Silk damping and slow growth of structure.  These effects (and others) are hard to survey with confidence, at
least for me; but I think they can only make a pretty good situation better.  Indeed, we know everything works out nicely for $r$ a little bigger than 1, so in that range we saturate the preceding estimate; 
these other complications will
mainly just suppress the competition.

\section{Implications}

The assumptions underlying the alternative axion cosmology I have pursued above have significant implications for axion physics, supersymmetry, and cosmology.  By following out these implications,
we might be able either to enhance the credibility of their application to describe reality, or to demolish that credibility.

For axion physics: Laboratory searches for solar axions within the ``astrophysical window'', or for cosmic background axions as dark matter, have been predicated on smaller values of $F$ than
what we assume here.   Large values of $F$ imply weaker coupling to matter, and render direct detection more difficult.   So, unfortunately, the anthropically interesting scenario is incompatible with
direct detection of axions in the foreseeable future.   Of course, I'd be quite happy to see it ruled out in this particular way!    On the other had, large values of $F$ would appear to have some theoretical advantages. 
It might be
possible to identify the Peccei-Quinn scale with the scale of gauge symmetry unification indicated by the successful running of couplings calculation, for example.   Independent of any particular model, 
the general idea that a single condensate might trigger breaking of several symmetries is quite attractive.   There is also some advantage to having inflation occur after Peccei-Quinn symmetry breaking, in that
axion strings, which certainly complicate and might ruin the cosmology of axion dark matter, get diluted away.  

For supersymmetry: If axions dominate the dark matter density, then of course the dark matter candidate that arises in many models of low-energy supersymmetry does not.   This candidate is often 
refereed to interchangeably as the WIMP (weakly interacting massive particle) or the LSP (lightest supersymmetric particle), but it be convenient here to make a distinction.  The framework in which the properties
of LSP/WIMP particles are discussed is most often, either explicitly or in effect, the minimal supersymmetric extension of the standard model.   In that framework, the lightest $R$-parity odd particle, the LSP,
is stable on cosmological timescales, and for an otherwise plausible range of parameters -- notoriously,
several of the phenomenologically crucial parameters in models of low-energy supersymmetry are at present poorly constrained -- 
one finds that
it is indeed a weakly interacting particle whose density is predicted, following out big bang cosmology, to be compatible with what the astronomers find for dark matter.   So the LSP can provide the cosmological
WIMP.   On the other hand, even within this framework there is an equally plausible range of parameters such the LSP is produced with too small a density to provide the cosmological WIMP.  
The scenario discussed above therefore favors that range.    

Along this line, if we accept the approximate equality of supersymmetric dark matter to baryonic matter as a {\it fait accompli\/} arising from a coincidence among disparate 
microscopic parameters, say for 
concreteness $\rho_{\rm LSP}/\rho_{\rm baryon} =3$, then our anthropic scenario would at least make the additional coincidence $\rho_{\rm axion}/\rho_{\rm baryon} =3$ appear less 
conspiratorial.   

Another possibility, which I find especially intriguing and not at all implausible, is that the lightest supersymmetric particle is {\it not\/} the partner of a standard model particle.  
It could be the gravitino, the dilatino, the axino, a modulino, a 
combination of these, ... .   In these cases the true LSP is generally a very feebly interacting particle, with coupling strength similar a graviton, axion, ... .  
The pseudo-LSP that will be observed (at the LHC, presumably)
as a standard model partner will decay into this true LSP.  
The decay will be rapid on cosmological timescales, so the pseudo-LSP sort of WIMP can't supply the cosmological dark matter.   
Since the true LSP is very feebly interacting and relatively light, direct production
of the true LSP during the big bang will not yield a cosmologically significant density of dark matter.   
It might be produced at a cosmologically significant level at relatively late times though decays of the pseudo-LSP, but it requires some special adjustments both to avoid wreaking 
cosmological havoc with these
decays and to reproduce the observed dark matter abundance.

It's at least equally plausible to suppose that the LSP is not produced enough to make the observed dark matter, and that is an important independent motivation 
to consider axions as an alternative.   An especially spectacular possibility is that the pseudo-LSP might electrically charged.  
Cosmologically stable charged matter in the form of mass $\sim 100$ GeV particles produced with cosmological density 
comparable to the observed dark matter density is a phenomenological disaster, but I am 
emphasizing that the pseudo-LSP need not be stable.  There are large, otherwise attractive regions of the parameter space for low-energy supersymmetry that have been excluded on these grounds, maybe prematurely.  
There is a wonderful signature for this possibility: the charged pseudo-LSP, produced at LHC, though unstable on cosmological timescales, could be stable on laboratory timescales.

For cosmology: The most distinctive features of axions as dark matter, to wit that they are produced cold, in fact so cold that they fill out a very small region of phase space and form caustics, continue to hold in
the alternative axion cosmology. 
If anything, their derivation is cleaner, since there are no axion strings, there is a clean specification of very simple initial conditions, and in post-inflation period, since temperatures are well 
below $F$, axions have only very feeble non-gravitational interactions.  

The initial misalignment angle, which eventually
materializes as the dark matter density, can provide an independent source of cosmological density perturbations, apart from ambient temperature fluctuations.   
If we take the inflationary origin of fluctuations at face value,
we find that this additional field provides a source of isocurvature fluctuations, whose amplitude depends on the scale of inflation.  
Recent observations put significant constraints on the amplitude of isocurvature fluctuations,
so the scale of inflation can't be too large; but perhaps the present scenario sharpens the motivation to search for them down to low levels \cite{krauss}.    

Finally, it is tempting to connect the line of thought pursued here with the other context in which anthropic reasoning has been applied to cosmology recently, that is Weinberg's discussion of the cosmological
term \cite{weinCosmo}.  He framed his discussion rather abstractly, without specifying a microscopic model.  It is quite simple to make a model along the lines discussed here.   
We can go back to the comforting -- but of course
totally unproved! -- assumption that
the asymptotic value of the cosmological term, in the distant future, is zero, and that what we are observing at present is residual energy 
frozen into a scalar axion-like field whose value is effectively uniform over the observable Universe.   This requires a small value of $\Lambda$ and a large value of $F$, relative to the axion that plays a role
in to strong P,T problem, in order that the mass $m\sim \Lambda^{2}/F$ should be of order the inverse Hubble time, to insure that the field is ``stuck''.   
$\Lambda \sim 10^{-12}$ GeV, $F\sim 10^{19}$ GeV
will do the job; these values also (barely!) assure, respectively, that the vacuum energy controlled by our field can supply enough for the observed cosmological term, 
and that it is associated with Planck-scale physics. 
The closeness of this call might be considered a small bit of encouragement for observational programs to check whether
the dark energy might have got unstuck, and started to evolve, in recent cosmological times.

It could appear highly unnatural, upon first sight, that symmetry breaking at such a large scale could be associated with so little energy; and in general it would be, but in axion physics it is not so unreasonable.   
The point is that all effects of $\theta$-like parameters are nonperturbative, and in weak coupling they contain explicit suppression 
factors like $e^{-8\pi^{2}/g^{2}}$.  In QCD that suppression is obscured, since $g$ is not uniformly small, 
but it doesn't take much smallness in the $g$ governing the relevant gauge theory to render this suppression factor quite small.   

Having made these alterations we can repeat our cosmological story, and -- within this circle of ideas -- 
justify Weinberg's hypotheses of effectively random variation of the cosmological term and its independence from other parameters.   
A minor difference is that negative values of the cosmological term do not appear.    It would be logical, of course, and very interesting, to consider from an anthropic perspective the implications of allowing both
the dark matter-baryon matter ratio $r$ and the cosmological term $\Lambda_{\rm C}$ to vary independently but simultaneously.    One must keep in mind that
the inflated PQ horizon might be quite different from the
corresponding horizon for the dark energy ``axion''.   If that occurs, then we should vary one mismatch angle over the 
Multiverse corresponding to the smaller horizon before varying the other over the MultiMultiverse associated with the larger horizon.   
The most probable value may be different, since the $y$ that maximizes
$f(x,y)$ is not the same as the $y$ that  maximizes the average $\langle f(x,y) \rangle_{x}$  taken over $x$, in general.  
A virtue of explicit dynamical models is that they bring subtleties like this into the foreground.


\begin{thebibliography}{99}
\bibitem{basic}
Here by ``basic'' I mean irreducible, and am consciously 
avoiding the freighted
term ``fundamental''.  No doubt there are 
extremely profound insights about how complex systems behave and develop to be derived  from the existence of mind, and in particular from its concrete emergence in history. 
Such insights will be fundamental by any reasonable standard, but not basic in the sense used here.
\end{thebibliography}
\end{document}